\documentclass[%
 aip,
 amsmath,amssymb,
 reprint,%
]{revtex4-1}
\usepackage{float}
\usepackage{graphicx}
\usepackage{dcolumn}
\usepackage{bm}

\usepackage[utf8]{inputenc}
\usepackage[T1]{fontenc}
\usepackage{mathptmx}
\begin{document}

\preprint{AIP/123-QED}

\title[]{On the merit of hot ion mode for tearing mode stabilization}

\author{S. Jin}
 \email{sjin@pppl.gov}

 \author{A. H. Reiman}%
 \email{areiman@pppl.gov}

\author{N. J. Fisch}
 \email{fisch@pppl.gov}
 
\affiliation{ Department of Astrophysical Sciences, Princeton University, Princeton, New Jersey 08543, USA}%

\affiliation{Princeton Plasma Physics Laboratory, Princeton, New Jersey 08540, USA}%

\date{\today}

\begin{abstract}
The stabilization of tearing modes with rf driven current benefits from the cooperative feedback loop between rf power deposition and electron temperature within the island. This effect, termed rf current condensation, can greatly enhance and localize current driven within magnetic islands. It has previously been shown that the condensation effect opens the possibility of passive stabilization with broad rf profiles, as would be typical of LHCD for steady state operation. Here we show that this self-healing effect can be dramatically amplified by operation in a hot ion mode, due to the additional electron heat source provided by the hotter ions. 

\end{abstract}

\maketitle

\section{Introduction}

The development of effective strategies for disruption avoidance is one of the primary objectives of magnetic fusion research. The heat and force loads to the machine walls caused by the sudden loss of confinement can cause unacceptable damage to the device\cite{boozer_2012}; this threat is only becoming more dangerous as tokamaks are designed with increasing levels of stored energy.\cite{allan} Large magnetic islands play a key role in triggering disruptions, thereby setting a primary performance limit in tokamaks.\cite{La_Haye_2006a,de_Vries_2014,de_Vries_2011}  Suppressing the growth of these unstable islands, known as neoclassical tearing modes (NTMs), is accordingly a crucial aspect of disruption avoidance.

Islands form at rational surfaces when helical magnetic perturbations disturb the typically toroidally nested magnetic topology. The resulting island flux surfaces enable rapid radial parallel transport, locally flattening the pressure profile over the extent of the island, in the absence of significant net heat sources or sinks. This creates a "hole" in the bootstrap current, which in turn, creates a magnetic perturbation that reinforces the initial helical disturbance, thereby driving the island unstable. Driving current via rf waves\cite{fisch_1987} in the island interior therefore suppresses island growth by effectively replacing the missing bootstrap current.\cite{Reiman_1983}.

The rf stabilization of tearing modes has been the subject of extensive theoretical and experimental work. Only recently, however, has the highly nonlinear nature of rf deposition in magnetic islands been identified. The closed magnetic topology of the island provides thermal insulation and reduces the cross-field thermal diffusivity.\cite{Spakman_2008,Inagaki_2004,Bardoczi_2016,Ida_2012} This produces significant electron temperature perturbations relative to the background plasma in the presence of rf heating, with reported relative temperature increases as large as 20\%.\cite{Westerhof_2007} Furthermore, the power that both lower hybrid (LH)\cite{fisch_1978} and electron cyclotron (EC) waves \cite{fisch_1980} can deposit is proportional to the number of resonant electrons ($P_{dep}\sim n_{res}$). This produces an exponential sensitivity to small temperature perturbations ($n_{res}\sim \exp(-w^2)\approx \exp(-w_0^2)\exp(w_0^2\Delta T_e/T_{e0})$), where $w:=v_{res}^2/v_{th}^2$ is the ratio of the resonant and thermal velocities. As $w_0^2\approx4-20$ in practice\cite{kfj}, even small increases in temperature can dramatically enhance the power absorption, and consequently, the stabilizing current driven. This current is also preferentially enhanced where it is most stabilizing, as the temperature profiles are governed by diffusion and typically centrally peaked. This amplification and focusing are termed the current condensation effect. \cite{Reiman_2018,Rodriguez_2019,eduardo,nies_2020,Jin_2020} 

As the nonlinear enhancement is dictated by the electron temperature perturbation ($P_{rf}\sim \exp(w_0^2 \Delta T_e/T_{e0})$, it naturally follows that additional heat sources to the electrons would further amplify the condensation effect and associated stabilization properties. Such a supplementary heat source could be provided by operating in a "hot ion mode", in which the ions are maintained at significantly higher temperatures than the electrons. The hot ion mode of operation has traditionally been of interest for its enhanced reactivity and improved confinement properties. Devices are typically limited by total plasma pressure, while the fusion cross-section depends only on the temperature of the fuel ions. Therefore, the electron temperature contributes to the shared pressure limit, without producing fusion power. Furthermore, hotter electrons incur greater power losses, through either radiation or transport. Experimentally, hot ion modes have exhibited dramatically enhanced confinement \cite{Petty2000}, and have achieved the top fusion power records to date.\cite{Bell1995,Keilhacker2001}

Yet, hot ion mode operation is no longer a priority in modern reactor design\cite{Sips2015,Chen2017}. This is largely due to the perceived constraint that $\alpha$-particles produced in burning plasmas will principally heat electrons. Since the $\alpha$-particle heating is the dominant heating mechanism in reactors, as opposed to present day devices where external heating sources can principally heat ions, there appears to be no way of achieving a hot ion mode in a reactor self-sustained by $\alpha$-particle heating, absent some kind of intervention. However, the $\alpha$-particle power may alternatively be diverted to rf waves by capitalizing on favorable population inversions along diffusion paths in energy-position space---an effect known as alpha channeling.\cite{Fisch1992,Fisch1999,Fisch_1994} In this way, much of the $\alpha$-particle power may ultimately by diverted to heating ions, supporting the ion temperature surplus. The synergy with disruption avoidance, identified here for the first time, significantly broadens the appeal of hot ion modes,  and calls for renewed attention to the subject.

Here we quantify the extent to which operating in a hot ion mode may facilitate the rf-stabilization of tearing modes. We show that even modest temperature differences between the ions and electrons can have significant impact on the rf power required and island width at stabilization, provided sufficient collisional energy coupling between electrons and ions. The rf-condensation effect plays an essential role---similarly dramatic hot-ion mode enhancements are not possible for other methods of current drive, e.g. ohmic, NBI. We also discuss the issues raised by the dual impact of energy coupling: although the enhancement to rf stabilization increases with coupling strength, so does the difficulty of maintaining a temperature differential. We show that accessing a synergistic regime for hot ion mode and tearing mode stabilization places constraints on coupling strength, power partitioning between species, and relative transport inside vs. outside of the island.

The paper is organized as followed. Section II introduces the two fluid equations for the island temperatures, and summarizes the essential features of the solutions. Section III characterizes the impact of the temperature differential and energy coupling on the rf condensation effect, in particular, the implications for self-healing in steady-state scenarios. Section IV provides a preliminary investigation of the accessibility of strongly synergistic regimes. Section V summarizes the main results and conclusions.

\section{Two-temperature island model}
The energy transport equations for electrons and ions can be written as:

\begin{equation}
\label{eq:general}
    \frac{3}{2} \partial_t n_s T_s-\nabla \cdot ( n_s \chi_s \cdot \nabla T_s)=\frac{3}{2\tau_{eq}} n_s (T_{r}-T_{s})+P_{s}
\end{equation}
where subscript $s$ denotes either electrons or ions; subscript $r$ denotes the other species; $\chi_s$ is the heat diffusivity tensor of species $s$; and $\tau_{eq}=\frac{m_i}{m_e} \frac{3}{8}\sqrt{\frac{m_e}{2\pi}}\frac{(kT_e)^{3/2}}{n e^4 \lambda} $ is the electron ion equilibration time, where $\lambda\approx 20$ is the Coulomb logarithm, $m_s$ is the mass of species $s$, $\mu$ is the ion mass in units of proton masses. $P_s$ contains whatever species specific heat sources and sinks may be present, aside from the explicitly written electron-ion equilibration term.

 Eqs. (\ref{eq:general}) can be greatly simplified due to the large disparity between the MHD time scales on which the island width and background plasma parameters evolve, and the characteristic diffusion times $\tau_{D,s}$ and energy equilibration rate $\tau_{eq}$. We may then consider "steady state" solutions for which $\partial_t\rightarrow0$, and background parameters are held constant, and linearize Eqs. (\ref{eq:general}) for perturbations $\widetilde{T_s}$ to the temperatures at the separatrix, $T_{s,0}$. In other words, at the separatrix, the electron temperature is $T_{e,0}$ and the ion temperature is $T_{i,0}$, where, in the hot ion mode $T_{i,0}>T_{e,0}$. On these timescales, the field lines will be approximately isothermal, so Eqs. (\ref{eq:general}) may be written as 1-D coupled diffusion equations for the perturbed temperatures $\widetilde{T_s}$. In slab geometry, the linearized Eqs. (\ref{eq:general}) become:
 \begin{equation}\label{eq:general2}
         \frac{3}{2} n_s \partial_t \widetilde{T_s}-n_s \chi_{s,\perp} \partial_x^2 \widetilde{T_s}=\frac{3}{2\tau_{eq}} n_s (\widetilde{T_r}-\widetilde{T_s}+T_{r,0}-T_{s,0})+P_{s}
 \end{equation}
 where $\chi_{s,\perp}$ is the perpendicular thermal diffusivity of species $s$ (hereafter we drop the $\perp$ subscript). Note the 0-th order heating term due to the temperature differential at the separatrix.  
 
For simplicity, we assume that the hot ion mode is sustained through ion heating sufficiently close to the core, rather than in the island, such that we may take $P_i=0$.  As for the electron source terms, it is typically found that pressure profiles flatten within the island in the absence of rf heating, and the usual condition of comparable electron and ion temperatures.\cite{Maraschek_2012} Then we might reasonably, and for the sake of simplicity, take $P_e=P_{rf}$.  Alternatively, even if the ohmic and radiation terms do not balance \cite{white_2015}, our selective choice of source terms is justified as long as they are negligible compared to the rf power. The nonlinearity of the rf power is retained in the following way:
\begin{equation}\label{eq:nlterm}
\begin{aligned}
    P_{rf}&\propto\exp(-w^2)=\exp(-m_e v_{res}^2/2(T_{e,0}+\widetilde{T_e}))\\ &\approx\exp(-w_0^2(1-\widetilde{T_e}/T_{e.0}))\\
    &\propto \exp(w_0^2 \widetilde{T_e}/T_{e,0})
\end{aligned}
\end{equation}
 
 Finally, Eqs. (\ref{eq:general2}) can be cast in a more illuminating form by scaling the spatial coordinate to the half island width ($x_{scl}:=W_i/2$) such that the island boundaries are at $x=\pm1$, and scaling the time coordinate to the electron diffusion time $\tau_{D,e}:=3 W_i^2/8\chi_e$. The nonlinear term $\ref{eq:nlterm}$ prescribes the scaled perturbed temperature $u_s:=w_0^2\widetilde{T_e}/T_{e,0}$. Multiplying Eqs. (\ref{eq:general2}) by $w_0^2 W_i^2/4nT_{e,0}\chi_e$ then yields:

\begin{equation}\label{eq:ue}
    -u_e''=P_0\exp(u_e)+c(u_i-u_e+h)
\end{equation}
\begin{equation}\label{eq:ui}
    -\gamma u_i''=c(u_e-u_i+h)
\end{equation}
subject to the boundary conditions $u_s(x=\pm1)=0$ at the island separatrix. Here $c:=\tau_{D,e}/\tau_{eq}$ is the ratio of the electron diffusion and electron-ion energy equilibration times, $h:= w_0^2(T_{i,0}-T_{e,0})/T_{e,0}$ is the normalized temperature differential at the separatrix, $\gamma:= \chi_{\perp,i}/\chi_{\perp,e}$ is the ratio of the electron and ion diffusivities, and $P_0$ is the rf power density normalized to $P_{scl}:=n T_{e,0}/w_0^2\tau_{D,e}$. We consider the case where the rf deposition is broader than the island width, as would be typical for LHCD, such that $P_0=const.$
\subsection{Naturally peaked electron temperature in hot ion mode}
Even in the absence of external electron heating $(P_0=0)$, magnetic islands in hot ion mode plasmas will exhibit peaked electron temperature profiles, given sufficient coupling $c$. Larger islands will have more strongly peaked electron temperatures, as $c\sim W_i^2$. Eqs.(\ref{eq:ue}) and (\ref{eq:ui}) admit the following solutions at $P_0=0$:
\begin{equation}\label{eq:esol}
    u_e(x)=\frac{h}{1+\gamma^{-1}}(1-\frac{\cosh(\sqrt{c(1+\gamma^{-1})}x)}{\cosh(\sqrt{c(1+\gamma^{-1})})})
\end{equation}

\begin{equation}\label{eq:isol}
    u_i(x)=\frac{h}{1+\gamma}(\frac{\cosh(\sqrt{c(1+\gamma^{-1})}x)}{\cosh(\sqrt{c(1+\gamma^{-1})})}-1)
\end{equation}
The electron temperature perturbation at the island O-point, $u_e(0)=h(1+\gamma^{-1})^{-1}(1-\cosh(\sqrt{c(1+\gamma^{-1})})^{-1})$, has the following limiting behaviors:
\begin{equation}
\begin{aligned}
        u_e(0) &\rightarrow hc/2 \qquad\qquad\qquad c\rightarrow0\\
        u_e(0) &\rightarrow h/(1+\gamma^{-1}) \qquad\quad c\rightarrow\infty
\end{aligned}
\end{equation}

In the low coupling limit ($c\rightarrow0$), the ion temperature profile is  minimally affected, so the electron heating is mostly determined by the temperature difference at the separatrix, $h$, and the ion diffusivity ratio, $\gamma$, does not enter. The temperature perturbation also increases linearly with coupling strength, i.e. quadratically in the island width ($c\sim W_i^2$). For strong coupling ($c\rightarrow\infty$), the electron and ion temperatures within the island equilibrate (except for a narrow boundary layer at $x=\pm 1$), and this balance is set by the diffusivities. A higher ion diffusivity $\gamma$ translates into a larger electron temperature, as a larger diffusivity means the ion temperature profile has greater resistance to being dragged down by the electrons. This can be most easily illustrated by the limit $\gamma\rightarrow\infty$: if the ions are extremely diffusive, their temperature profile remains flat, while the strong coupling pulls the electron temperature up to the separatrix temperature differential $h$.

A peaked electron temperature in hot ion mode islands is then a simple consequence of the bulk temperature differential $h$, and the disparate power balance within vs. outside of the island. Within the island, the strength of coupling is set not only by the equilibration rate, but depends on the ratio of equilibration and diffusion times ($c:=\tau_{D,e}/\tau_{eq}$). In the larger plasma, especially closer to the core, the temperature differential will be determined by a more complicated balance of heating and transport processes. The degree of equilibration can therefore be quite different inside vs. outside of the island, such that a large temperature differential $h$ may be supported at the separatrix, while the electron and ion temperatures within the island equilibrate more effectively in comparison, producing peaked electron temperatures. This point is discussed in more detail in Section IV. Although this temperature perturbation alone can have stabilizing properties, as any Ohmic current will concentrate to some extent near the island center (as the resistivity scales as $\sim T_e^{-3/2}$), the full potential of this hot ion mode induced electron temperature peaking can be best realized in combination with rf power deposition.

\subsection{Amplified rf condensation effect}

The current condensation effect enters through the nonlinear rf heating term, $P_{0}\exp(u_e)$. Following the treatment in Ref. \onlinecite{Reiman_2018}, we take $P_0$ (the power deposition profile in the absence of any temperature perturbation) to be a constant. This "power bath model" is appropriate for rf deposition profiles that are wide compared to the island, as would be typical of LHCD, and in sufficiently high power regimes such that the rf would not be depleted within the island\cite{fisch_1987}.

\begin{figure}[h]
\centering     
\includegraphics[width=\linewidth]{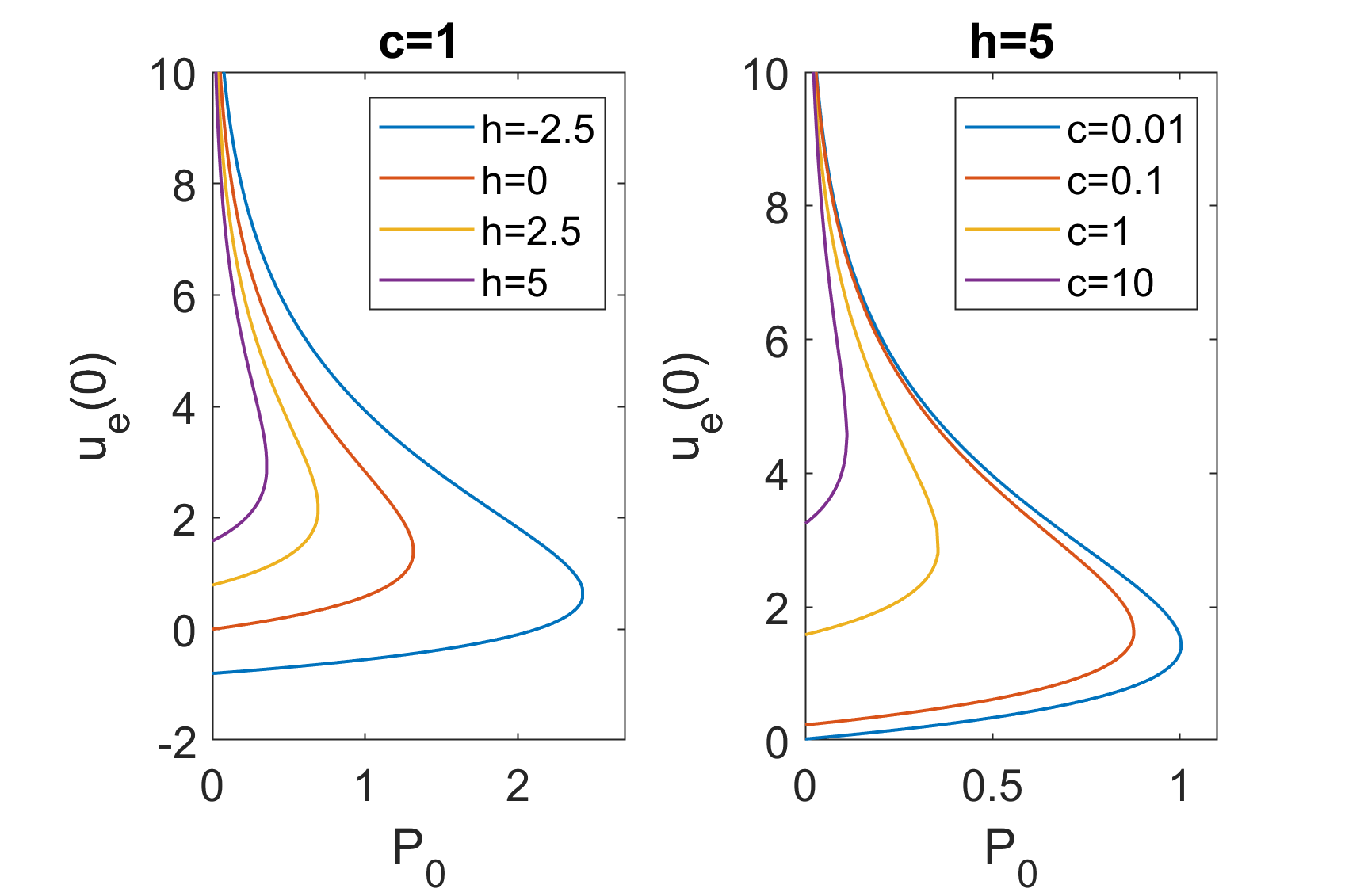}
\caption{\label{fig:ues}Electron temperatures vs scaled power $P_0$, for $\gamma=2$, $w_0^2=10$.}
\end{figure}

Eqs. (\ref{eq:ue}) and (\ref{eq:ui}) admit two branches of solutions, with the stable lower and unstable upper branches joined at a bifurcation point (technically a surface in the $c-h-P_0$ parameter space, but we use the language "bifurcation point" to refer to the $P_0$ at bifurcation for a given $c,h$), as shown in Fig. \ref{fig:ues}. At rf powers past the bifurcation point, the temperature in the island will continue to grow until encountering additional physics not included here, such as stiffness $\cite{eduardo}$ or ray depletion $\cite{Rodriguez_2019,Jin_2020,nies_2020}$. We opt for the simplified rf model used here to isolate the physics of the condensation effect in hot ion mode, without introducing particular depletion mechanisms; in exchange, the present investigation is limited to weak absorption regimes, i.e. the lower branch shown in Fig. 1. 

The $h=0$ case reduces to the two-fluid equations studied in Ref. \onlinecite{jin_2021}, the basic properties of which can be briefly summarized as follows. As the coupling strength $c\rightarrow0$, the electron and ion temperatures decouple, leaving a single-fluid equation for the electron temperature:$-u''=P_0\exp{(u)}$, while the ions remain unheated. As $c\rightarrow\infty$, the electrons and ions fully equilibrate ($u_i\rightarrow u_e$), again leaving a single-fluid equation for the shared temperature, except with the effective power reduced from the electron-only case by a factor of $1+\gamma$ due to the additional diffusive losses incurred through the ions: $-u''=P_0(1+\gamma)^{-1}\exp{(u)}$. Increased coupling always reduces the efficacy of electron heating for equal temperatures at the separatrix ($h=0$).

\begin{figure}[h]
\centering     
\includegraphics[width=\linewidth]{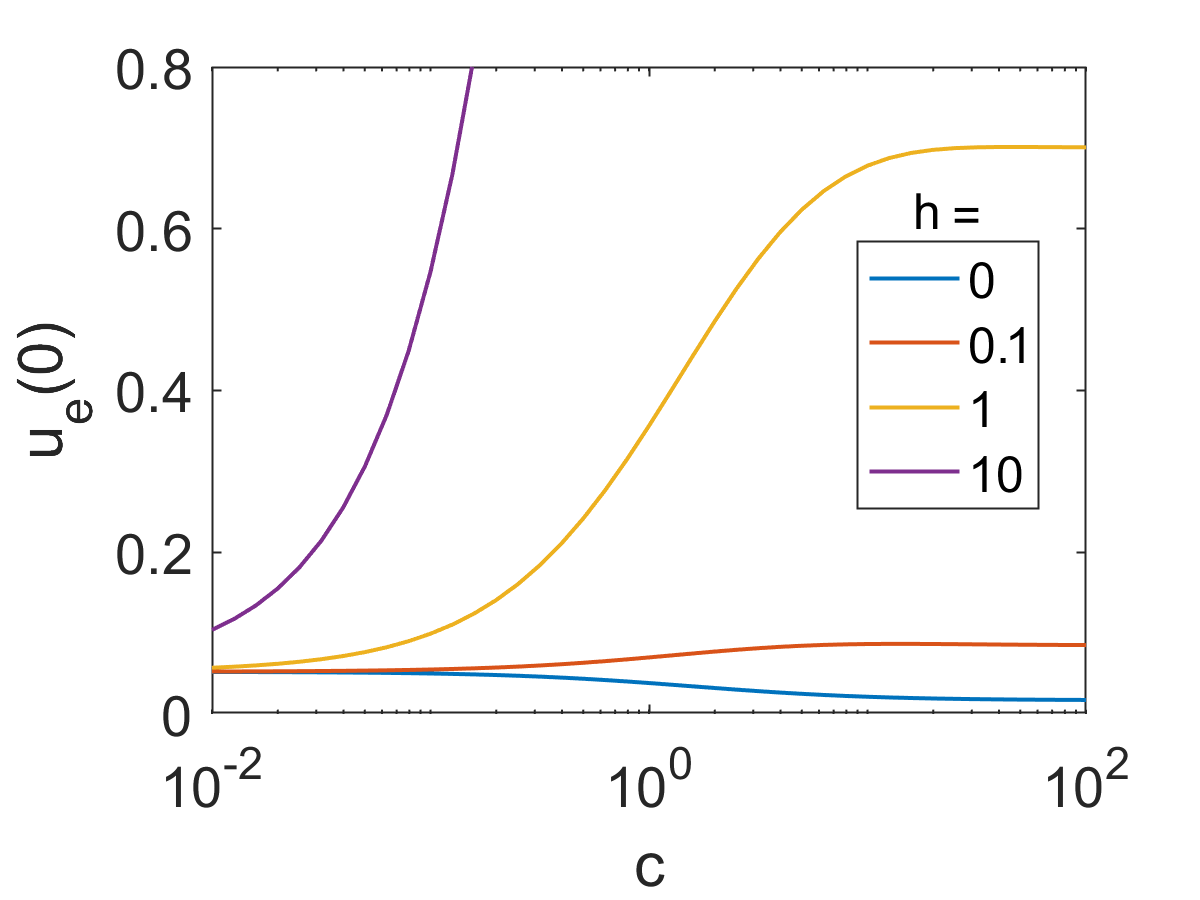}
\caption{\label{fig:uvc}Electron temperatures vs coupling strength $c$, for $\gamma=2$, $w_0^2=10$, $P_0=0.1$.}
\end{figure}

In contrast, energy coupling with the ions can provide net electron heating in a hot ion mode ($h>0$), such that the electron temperature increases with coupling strength $c$, as shown in Fig. \ref{fig:uvc}. Net electron heating occurs with increasing coupling $c$ for a given $P_0, \;h$ as long as $\int_{-1}^1 dx\; u_{e,c=0}(x)<2h$. Even if the temperature differential $h$ does not meet the criterion for net heating relative to the $c=0$ case, any $h>0$ still corresponds to a higher electron temperature than the $h=0$ case at a fixed coupling strength $c$. Since the rf absorption is exponentially sensitive to the electron temperature $u_e$, even small temperature differentials ($h$) are amplified through the condensation effect and can therefore significantly increase total heating, provided sufficient coupling.

Since larger electron temperatures may be achieved for the same unperturbed rf power $P_0$, and the absorbed power is given by $P_0\exp(u_e)$, this amounts to an even more dramatic enhancement of the driven current. Hot ion mode operation $(h>0)$ would therefore allow for islands to be stabilized at smaller widths, for a given rf power.  Alternatively, if the goal is for islands not to surpass a certain target width, it could cost much less rf power to accomplish this goal. The extent to which stabilization outcomes are affected by the hot ion mode enhanced condensation effect, in terms of power saved or width at stabilization, is quantitatively established in the following Section III.

\section{Enhanced stabilization in hot ion mode}
The rf condensation effect supports a self-healing property in tokamaks with a significant portion of the toroidal current driven by LHWs.\cite{jin_2021} As an island grows in the presence of ambient rf power, the electron temperature perturbation and resulting stabilizing current increase with island width. This is due both to the increasing total absorbed power $P_{tot}\sim W_i$, as well as the increasing confinement time $\tau_{D,e}\sim W_i^2$. In the previously studied $h=0$ case\cite{jin_2021}, even though parasitic heat losses to the ions also increase with island width ($c\sim W_i^2$), the former two effects dominate. As a result, islands may be passively stabilized by utilizing the ambient LHCD used for toroidal current drive, in combination with the natural tendency of the island to heat up as it grows; no external intervention\cite{Volpe_2009,Nelson_2019,fu_2020} (e.g. modulation, steering) would be required. The extent to which this self-healing mechanism is improved by hot ion mode operation can be calculated as follows.

The width at stabilization can be calculated with the modified Rutherford equation (MRE), which gives the island growth rate as a function of various driving and stabilizing mechanisms\cite{De_Lazzari_2009,Bertelli_2011}:
\begin{equation}
    \frac{\text{d}w}{\text{d}t}\propto \Delta_0'(w)-\Sigma_i\Delta_i'(\delta j_i)
\end{equation}
where $\Delta_0'$ is the classical stability index, and $\Delta_i'(\delta j_i)$ are corrections to the classical tearing mode equation due to perturbations of the parallel current at the resonant surface. These corrections have the form \cite{De_Lazzari_2009}:

\begin{equation}
\begin{aligned}
        \Delta_i & \propto \int_{-\infty}^{\infty}\text{d}x\oint\text{d}\xi \cos(\xi)\delta j_i (x, \xi)\\
         & =\int_{-1}^{1}\text{d}\Omega\int_{-\hat{\xi}}^{\hat{\xi}}\text{d}\xi \frac{\cos(\xi)}{\sqrt{\Omega+\cos(\xi)}}\delta j_i (\Omega, \xi)\\
         & =\int_{-1}^{1}\text{d}\Omega F(\Omega)\overline{\delta j_i (\Omega)}
\end{aligned}
\end{equation}
where $\hat{\xi}=\cos^{-1}(-\Omega)$, $x:=r-r_s$ is the radial displacement from the resonant surface, $\Omega$ is the flux surface label and $\xi:=:=\theta-\frac{n}{m}\phi$ is the helical phase, where $\theta$ ($\phi$) and $m$ ($n$) are the poloidal (toroidal) angle and mode number respectively. $\overline{\delta j_i(\Omega)}$ denotes the flux surface averaged perturbed current, and $F(\Omega):=\int_{-\hat{\xi}}^{\hat{\xi}}\text{d}\xi \frac{\cos(\xi)}{\sqrt{\Omega+\cos(\xi)}}$ serves as a flux surface weighting function.

The dominant driving term for NTMs will be due to the perturbed bootstrap current $\delta j_{bs}$\cite{Westerhof_2016}, while the stabilizing term of interest for our purposes comes from the perturbed LH current $\delta j_{LH}$. We may then take as the stabilization condition: $0\approx \Delta'_{bs}+\Delta'_{LH}$. Following common practice\cite{Bertelli_2011}, we take the perturbed bootstrap current to be constant over the island, such that $\delta_{bs}=-J_{bs}$, where $J_{bs}$ is the bootstrap current density at the resonant surface prior to the island formation. The corresponding contribution to the island growth rate is then:
\begin{equation}
    \Delta'_{bs}\propto -J_{bs}\int_{-1}^{1}\text{d}\Omega F(\Omega)
\end{equation}

The perturbed rf current $\delta_{LH}$ is given by the nonlinear enhancement to the LH current at the resonant surface, due to the the heating of the island. The current driven is roughly proportional to the power deposition, neglecting the dependence of the current drive efficiency on temperature, which is negligible compared to the exponential enhancement factor\cite{fisch_1987}. Then, the rf contribution to the island growth rate is:
\begin{equation}
    \Delta'_{LH}\propto J_{LH}\int_{-1}^1\text{d}\Omega (\exp(u_e(\Omega))-1)F(\Omega)
\end{equation}
where $J_{LH}$ is the LH current at the resonant surface prior to the island formation. 

In order to obtain the island temperature in terms of the flux function $\Omega$, we now write Eqs. (\ref{eq:ue})-(\ref{eq:ui}) in island geometry\cite{Reiman_2018}:
\begin{equation}\label{eq:ueig}
    \hat{D}u_e=P_0\exp(u_e)+c(u_i-u_e+h)
\end{equation}
\begin{equation}\label{eq:uiig}
    \gamma\hat{D}u_i=c(u_e-u_i-h)
\end{equation}
where $\hat{D}$ is a diffusion operator that accounts for the geometry of the flux surfaces within the island:
\begin{equation}
    \hat{D}:=-\frac{1}{\rho K(\rho)}\frac{d}{d\rho}\;\frac{E(\rho)-(1-\rho^2)K(\rho)}{\rho} \;\frac{d}{d\rho}
\end{equation}
where $K$ ($E$) is the complete elliptic integral of the first (second) kind, and $\rho$ is an alternate flux coordinate such that $\Omega:=2\rho^2-1$, and $\rho=0\;(1)$ at the island center (separatrix).

The stabilization condition follows by balancing the contributions of the LH and bootstrap currents:
\begin{equation}\label{eq:stabcond}
    1=R^{-1}\frac{\int_{-1}^1\text{d}\Omega (\exp(u_e(\Omega))-1)F(\Omega)}{\int_{-1}^{1}\text{d}\Omega F(\Omega)}
\end{equation}
where $R:=J_{bs}/J_{LH}$ is the ratio of the bootstrap to LH currents prior to island formation. The island width and temperature differential enter the stabilization condition Eq. (\ref{eq:stabcond}) through $u_e$, via the parameters $P_0$, $c$, and $h$ respectively.

\begin{figure}[h]
\centering     
\includegraphics[width=\linewidth]{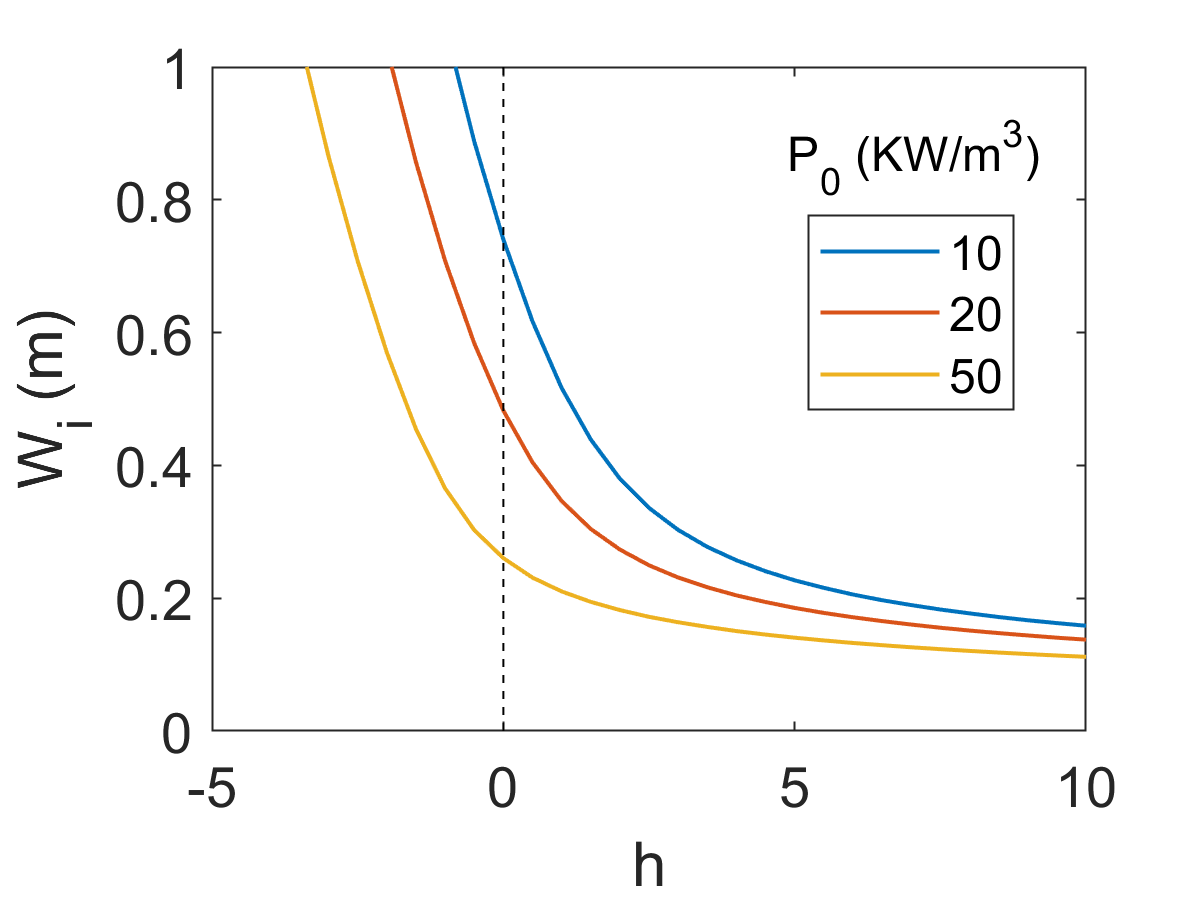}
\caption{\label{fig:wstab}Island width at stabilization vs. temperature differential $h$ for several rf powers, calculated with $\gamma=2$, $R=1$ and plasma parameters of $n=10^{20} \,m^{-3},\; T_{e0}=5\,keV$, $\chi_e=0.1 m^2/s$. As a reference, $c=1$ at $W_i\approx0.2 \,m$.}
\end{figure}

Figure \ref{fig:wstab} illustrates how the temperature differential $h$ affects the width at which the stabilizing rf contribution to the MRE overtakes the destabilizing bootstrap term, for a 50-50 rf-boostrap current split ($R=1$). Of course, relatively larger bootstrap currents $R>1$ require larger rf powers/larger island widths to be stabilized; this dependence has been explored in Ref. \onlinecite{jin_2021}, and is thus not repeated in detail here. Plasma/rf parameters are taken from previous simulations performed for ITER scenario 2\cite{Frank_2020}, except with various ion temperature differentials assumed but not self-consistently calculated. The caveats of this approach are discussed in Section IV.

It can be seen that hot ion mode operation ($h\approx2.5-10$), would allow for islands to be stabilized at dramatically smaller widths for the same rf power. For a local LH power density of $10 \;KW/m^3$ with $h=0$, islands would self stabilize only after reaching an unacceptably large size of $\sim0.8\;m$. The stabilization width can be halved by an ion temperature surplus of $h=2.5$ ($25\%$ temperature difference), or even quartered by an ion temperature surplus of $h\approx5$. Fig. \ref{fig:wstab} can also be interpreted in terms of power savings for a given width at stabilization. Stabilizing islands at $\sim25 \;cm$ would require $\sim 50\; KW/m^3$ of rf power, in the absence of an ion temperature surplus. This can be achieved for roughly half the power if $h\approx2.5$, or a fifth of the power if $h\approx5$.

Note that a given ion temperature surplus makes a larger difference for smaller rf powers. This is most obviously, but incompletely, explained by the simple fact that the relative power contribution of the ions ($ch/P_0$) is larger for smaller rf powers. Additionally, smaller rf powers allow the island to grow to larger widths, which strengthens coupling ($c\sim W_i^2$) and the corresponding electron heating for a given temperature differential, $ch$. For similar reasons, there are also diminishing returns with increasing $h$. Once the hot ion mode enhanced condensation effect is already suppressing the island to widths where coupling is negligible, additional increase of $h$ will no longer effectively provide additional heating. However, it happens to be the case that at the experimentally relevant plasma and rf parameters used in Fig. \ref{fig:wstab}, the coupling is sufficiently strong such that $h=0$ is well before the point of diminishing returns, i.e. hot ion mode significantly improves stabilization outcomes.

\begin{figure}[h]
\centering     
\includegraphics[width=\linewidth]{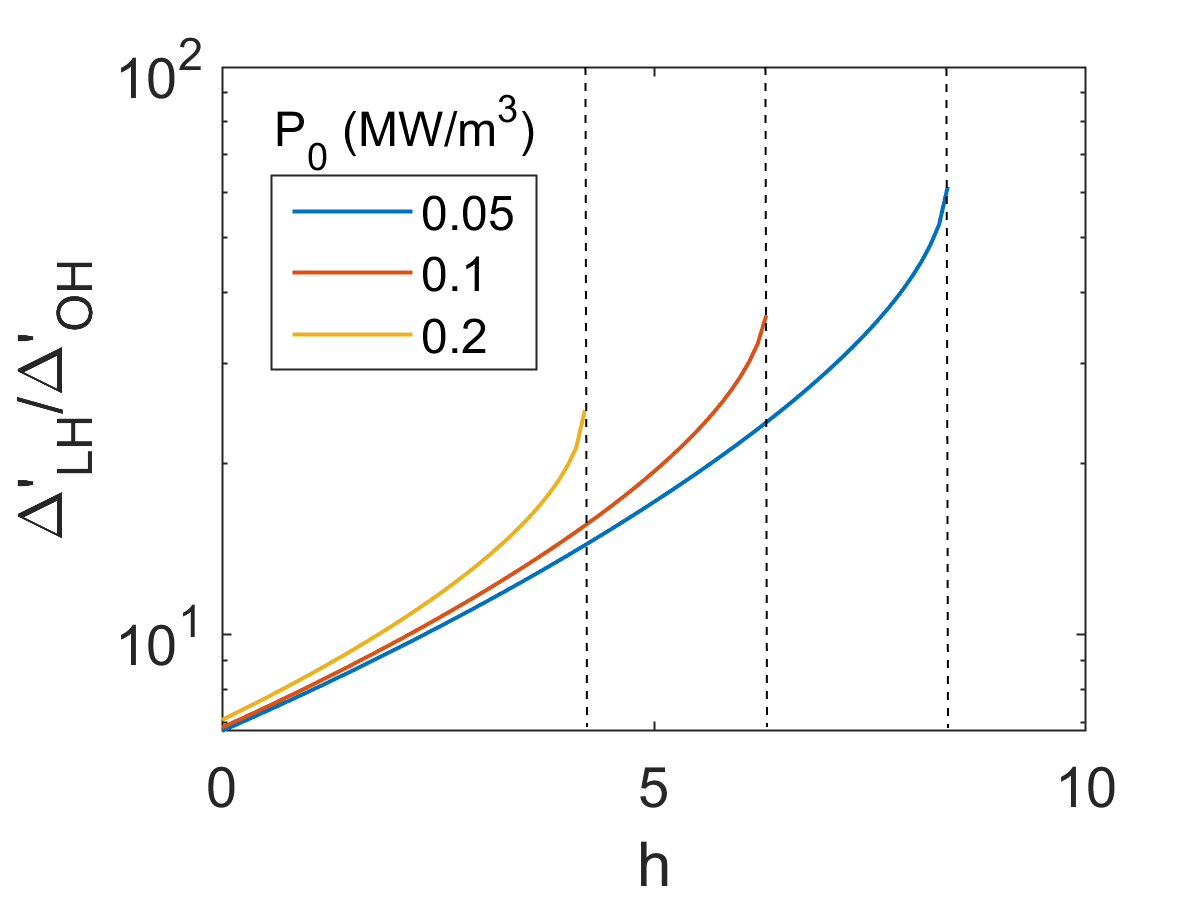}
\caption{\label{fig:comp}Relative contributions to MRE of LHCD and inductive currents ($\Delta'_{LH}/\Delta'_{OH}$) vs. temperature differential $h$ for several rf powers, calculated with $W_i=20 \,cm$, $\gamma=2$, $n=10^{20} \,m^{-3},\; T_{e0}=5\,keV$, $\chi_e=0.1 m^2/s$. Dashed lines indicate bifurcation thresholds.}
\end{figure}

It is worth emphasizing that the potency of this hot ion mode stabilization enhancement is only made possible by the nonlinearity of rf deposition. The exponential enhancement of driven current is unique to rf waves that act on the tail of the electron distribution. Consider, for comparison, inductively driven current. While ohmic current is also enhanced by the electron temperature, the form of the perturbed current is $\Delta j_{OH} = \frac{3}{2} \Delta T_e/T_{e0}$. Fig. \ref{fig:comp} shows the relative contributions to the island growth rate of rf and inductive currents of equal unperturbed magnitude. Not only is the rf contribution larger even in the absence of an ion temperature surplus (to be expected from rf condensation alone), the relative stabilization contribution grows super-exponentially with $h$. Evidently, LHCD and hot ion mode operation are distinctly synergistic for tearing mode stabilization.
\section{Accessibility of Synergistic regimes}

As far as tearing mode stabilization is concerned, the stronger the coupling to hotter ions, the better. However, from the perspective of hot ion mode operation, increased coupling makes it harder to maintain the temperature differential. 
The question at hand is then, how feasible is a regime in which the effective coupling outside the island is weak enough to support a substantial ($h\approx 5-10$) temperature differential in the bulk plasma, while the coupling within the island remains strong enough for there to be significant electron heating from the hotter ions? In other words, is the region in $c-h$ parameter space where rf condensation is greatly enhanced consistent with hot ion mode operation?

\begin{figure}[h]
\centering     
\includegraphics[width=\linewidth]{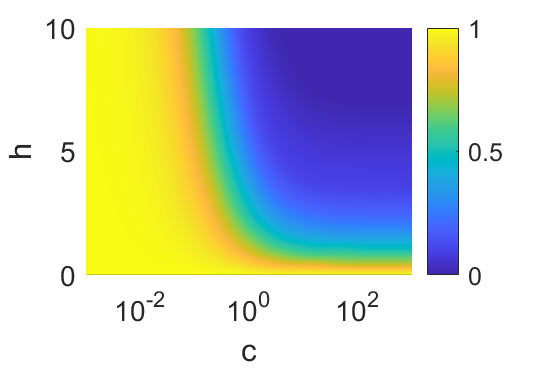}
\caption{\label{fig:pbifred}Reduction of $P_{bif}$ from ion temperature surplus $h$ for fixed $c$.}
\end{figure}

First, the requirements for "substantial" synergy must be explicitly stated in terms of $c$ and $h$. Although the nonlinear effects from current condensation are present at lower powers and temperatures as well, the bifurcation point provides an unambiguous marker for the strongly nonlinear regime. As previously discussed, past the bifurcation point, the island temperature will dramatically increase until it is ultimately limited by an additional loss mechanism such as stiffness \cite{eduardo} or rf depletion\cite{Rodriguez_2019,Jin_2020}. The bifurcation point must also be reached in order to access a hysteresis effect\cite{Reiman_2018} that could allow for further stabilization efficiency. The impact of hot ion mode on the condensation effect can therefore be simply quantified by the reduction in the power at bifurcation ($P_{bif}$) achieved by a temperature differential $h$, at a given coupling strength $c$ (Fig. \ref{fig:pbifred}).

Hot ion modes with dramatically improved reactivity and confinement properties have $T_{i,0}/T_{e,0} \approx 1.25\;-\;2$\cite{Fisch_1994,Clarke1980} or $h \approx 2.5\;-\;10$. For these temperature differentials, it can be seen from Fig. \ref{fig:pbifred} that the accessibility of the bifurcation is substantially improved ($P_{bif}$ halved) for coupling strengths $c\sim\mathcal{O}(1)$. Having established the $c$ and $h$ values corresponding to strong synergies between hot ion mode and rf stabilization of islands, they must be checked for consistency with overall power balance constraints. While a fully self consistent calculation would fall well beyond the scope of this paper, here we will outline the main aspects of the problem, and in doing so, formulate conditions for accessing such a synergistic regime, and at least roughly constrain the accessible c-h parameter space.

In principle, the parameter $h$ that enters Eqs. (\ref{eq:ue}-\ref{eq:ui}) is a local quantity, and would depend on the $q$, $T_e$ and $T_i$ profiles. Modelling such profiles presents a formidable challenge\cite{Candy2009,Lopez2018,emdee2021}, even in the absence of off-normal events such as island formation. We therefore opt for simplicity at the expense of precision with the following 0-d power balance model\cite{Fisch_1994}:
\begin{equation}\label{eq:powbal}
    \frac{d U_s}{dt}=-U_s/\tau_s+(U_r-U_s)/\tau_{eq}+P_s
\end{equation}
where $U_s:=\frac{3}{2}nT_s$ is the energy density of species $s$ and subscript $r\neq s$ indicates the other species, $P_s$ is the external heating for species $s$, $\tau_{eq}$ is the electron-ion energy equilibration time as defined in section II, and $\tau_s$ is the energy confinement time of species $s$ which is an aggregate measure of various heat loss processes, e.g. conduction, convection, radiation. Restricting our attention to steady state solutions, and rescaling Eqs. (\ref{eq:powbal}) gives the following:
\begin{equation}\label{eq:powbe}
    0=-1+c_0 h_0+(1-\eta)P
\end{equation}
\begin{equation}\label{eq:powbi}
    0=-\gamma_0(h_0+1)-c_0 h_0 +\eta P
\end{equation}
where $\eta:=P_i/(P_e+P_i)$ is the fraction of total heating that goes to the ions, $P:=(P_e+P_i)/(U_e/\tau_e)$ is the scaled total heating power, $c_0:=\tau_e/\tau_{eq}$ is the ratio of the electron confinement to the energy equilibration times, $h_0:=(U_i-U_e)/U_e$ is the temperature differential, and $\gamma_0:=\tau_e/\tau_i$ is the ratio of electron to ion confinement times. $c_0$
, and $\gamma_0$ are defined analogously to their counterparts within the island, but note that $h$ is defined with an additional factor of $w_0^2\approx10$ compared to $h_0$. In general, $\gamma_0\neq \gamma$, $h_0\neq h$ and $c_0\neq c$.  Eqs. (\ref{eq:powbe}-\ref{eq:powbi}) can be converted in to a relationship between $\eta$, $\gamma_0$, $c_0$, and $h_0$:
\begin{equation}\label{eq:hc0}
    h_0=\frac{-\gamma_0(1-\eta)+\eta}{c_0+\gamma_0(1-\eta)}
\end{equation}
In this form, some of the basic intuitions for hot ion mode are readily available: that hot ion modes are harder to sustain with stronger coupling ($h_0\rightarrow0$ as $c_0\rightarrow\infty$), and that the ion temperature surplus increases with the ion heating fraction ($dh_0/d\eta>0$) and decreases with the ratio of confinement times ($dh_0/d\gamma_0<0$). It is also immediately apparent from the numerator of Eq. (\ref{eq:hc0}) that hotter ions are possible only if: $\gamma_0<\eta/(1-\eta)$. To get a sense of what this means in terms of reactor-relevant parameters\cite{Fisch_1994,Xu2016}, consider $n=10^{20}\,m^3$, $T_e=10 \,keV$, and $\tau_e=0.1 \,s$, such that $c_0\approx0.3$. Then to achieve ions that are twice as hot as the electrons ($h_0=1$), this requires an ion heating fraction of $\sim 75\%$ if $\gamma_0=1$, or $\sim 65\%$ if $\gamma_0=0.5$.

It is important to note that the total 0D heating fraction $\eta$ and ratio of confinement times $\gamma_0$ are effectively independent. Although in Eqs. \ref{eq:ue}-\ref{eq:ui} for the island temperatures, we assume that the external heating to the ions is neglible, this is only a local assumption at the island flux surfaces, and is entirely compatible with substantial ion heating concentrated near the core of the plasma. As for the confinement time ratios, the heat transport in the island is predominantly diffusive so $\gamma$ reduces to the ratio of diffusivities, $\chi_i/\chi_e$---in the case of turbulent transport $\gamma\approx2$ while if turbulence is suppressed such that transport is nearly neoclassical, $\gamma\approx10$. A larger $\gamma$ supports a larger electron temperature perturbation in a hot ion mode (see Eq. \ref{eq:esol}). For the 0D model of the larger tokamak, $\gamma_0$ must encapsulate contributions from a variety of heat transport processes, e.g. convection, radiation, internal transport barriers (ITBs), etc. For the purposes of maintaining the ion temperature surplus, a smaller $\gamma_0$ is desirable. As a result of this independence between $\gamma$ and $\gamma_0$, and $\eta$ and our power balance assumptions within the island, we treat $\gamma_0$ and $\eta$ as parameters that are to be constrained by the synergy requirements of hot ion mode and NTM stabilization.

What remains is to establish a relationship between $c_0$ and $h_0$ and their counterparts within the island, c and h. To the extent that the bulk 0D temperature difference $(U_e-U_i)/U_e$ would be representative of its value at the island resonant surface, we can take $h\approx w_0^2 h_0$. Relating $c$ to $c_0$ is not as simple. As a first step, again assuming bulk quantities are sufficiently representative of their value at the island, we can take $tau_{eq}$ to be the same in $c_0=\tau_e/\tau_{eq}$ and $c=\tau_{D,e}/\tau_{eq}$. Then, the ratio $r_c:=c/c_0=\tau_{D,e}/\tau_e$ encapsulates the primary source of uncertainty in relating the electron heat transport inside vs. outside of the island. Then, in terms of the free parameter $r_c$, Eq. (\ref{eq:hc0}) can finally be written as a relationship between $c$ and $h$, which together determine the strength of the hot ion mode enhancements of rf stabilization:

\begin{equation}\label{eq:hc}
    h\approx w_0^2\frac{-\gamma_0(1-\eta)+\eta}{c/r_c+\gamma_0(1-\eta)}
\end{equation}

We first consider the case where $\eta=1$, assuming the electron heating from LHCD is a small fraction of the ion heating. In this case, Eq. \ref{eq:hc} reduces to the following simple form:
\begin{equation}\label{eq:eta1}
    h c\approx w_0^2 r_c\propto W_i^2
    \end{equation}

As previously mentioned, $\tau_e$ is an aggregate measure of heat loss processes, including conduction. Smaller $\tau_e$, with all else being equal, would raise $\gamma_0$, thus making the maintenance of an ion temperature surplus more difficult. We may then obtain conservative estimates for the strength of the hot ion mode/rf stabilization synergy by adopting a lower bound on $r_c$ by neglecting other loss processes, e.g. radiation and convection. In this case $r_c=\tau_{D,e}/\tau_e\approx \frac{W_i^2}{a^2}\frac{\chi_{e,0}}{\chi_e}$, where $a$ is the plasma minor radius, and $\chi_{e,0}$ is the bulk electron perpendicular thermal diffusivity. Thermal diffusivities within the island are reduced from those outside the island by 1-2 orders of magnitude\cite{Spakman_2008,Bardoczi_2016,Ida_2012,Inagaki_2004} (again, assuming the bulk value $\chi_{e,0}$ is comparable to the separatrix value). We then take $r_c\approx 10 (W_i/a)^2$, with $w_0^2=10$, giving us the delightfully simple expression, $hc\approx 100 (W_i/a)^2$. Then, for example, a hot ion mode with $h\approx 5$, the desired $c\sim 1$ can then be reached with manageable fractional island widths of $W_i/a\approx 0.2$.

Relaxing the ion heating fraction reintroduces the dependence on the confinement time ratio $\gamma_0$ (Eq. \ref{eq:hc}). Once again taking $h\approx5$ as a target operating point for hot ion mode, Eq. \ref{eq:hc} then prescribes a requisite ion heating fraction for a given confinement ratio $\gamma_0$. If $\gamma_0=1$ for example, adopting the previous conservative assumptions regarding the confinement time ratio such that $r_c\approx10 (W_i/a)^2$, the desired $c\sim1$ coupling regime can be achieved for a fractional island width of $W_i/a\approx0.3$ if $\eta=0.75$, but will require an unacceptably large $W_i/a\approx0.5$ if $\eta=0.55$. 

It is worth noting, however, that the above constraints on operating parameters (Eq. \ref{eq:hc}) are based on a somewhat artificially strict definition of "substantial" synergy (halving of the bifurcation power). An unfortunate consequence of quantifying the "required" conditions is the implication of a false binary, i.e. a regime is either synergistic or not. The benefits of hot ion mode for rf stabilization will certainly be present, if not as dramatic, for a much broader range of the $c-h$ parameter space than the $c\gtrapprox1$, $h\gtrapprox5$ target explored in this section. That such stringent goals can still be met for a range of operational parameters $\eta$ and $\gamma_0$, while adopting conservative assumptions regarding bulk confinement times (e.g. no radiation, convection), strongly indicates the accessibility of the synergistic regime for hot ion mode and rf stabilization.

\section{Summary}
The utility of hot ion mode operation extends beyond the traditionally recognized fusion power gains, to the synergistic stabilization of tearing modes with LHCD. Due to the nonlinear enhancement of LHCD with small temperature perturbations within the island, further electron heat sources can dramatically amplify stabilization via the rf condensation effect. Thus, in a hot ion mode, islands may be stabilized at smaller widths for the same rf power, or require less power to be stabilized at a given island size. Significant synergistic enhancement requires not only an ion temperature surplus, but also sufficient energy coupling within the island, such that $c,\,h\gtrsim1$.

While the effective heat source to the electrons increases with coupling for a given ion temperature surplus, this temperature differential will be difficult to maintain if coupling is too strong. Getting the maximum stabilization from hot ion mode operation is therefore not nearly as simple as "the hotter the better", and "the more coupled the better", as the temperature differential and coupling cannot be compatibly and simultaneously maximized. Despite approximations, our preliminary investigation of the accessibility of strongly synergistic $c-h$ regimes reveals the major issues at play, i.e. the importance of the energy confinement times of island vs. the bulk plasma, and requirements on operational parameters such as heating fraction $\eta$ and relative energy confinement of the ions $\gamma_0$. Contingent upon the feasibility of sufficiently high ion heating (large $\eta$) or preferential ion energy confinement (small $\gamma_0$) from an engineering perspective, the strongly synergistic $c-h$ regime is indeed accessible.

We do not address exactly how the hot ion mode is attained; we  only assume that somehow it is attained.  Attaining a hot-ion mode in a reactor, however, where the main heating is necessarily through the fusion reaction, requires some form of $\alpha$-channeling, in which the energy from fusion byproducts is channeled into a wave (avoiding collisional heating of the electrons), and that wave deposits its energy into the fuel ions  \cite{Fisch1992}.  
The attainment of a hot ion mode by $\alpha$-channeling is admittedly speculative, but with high upside potential. 
The main wave candidates are generally in the ion cyclotron range of frequencies, although combinations of more than one wave will probably give the optimal results \cite{Valeo1994,Fisch1995ii,Fisch1995,Chen2016a,Gorelenkov2016,Cook2017,Cianfrani2018,Cianfrani2019,Castaldo2019,Romanelli2020,White2021}.  
In all of these cases, the hot-ion mode requires significant wave intervention that first captures the $\alpha$-particle energy energy in the form of a wave which then damps on fuel ions. For example, waves amplified by $\alpha$-channeling might damp on the tritium ions directly as they encounter the tritium resonance \cite{Valeo1994}.  An alternative strategy would be to extract the $\alpha$-particle charge by waves, so that the energy then resides in plasma rotation, which might then be arrange such that the viscous dissipation of plasma rotation would favor ion heating rather than electron heating \cite{kolmes2021natural}.
For the purposes here, it does not matter which means exactly produces the hot ion mode.  So long as the diversion of $\alpha$-particle to fuel ions occurs outside of the island region, the boundary conditions on the electron and ion temperatures at the island periphery will simply be, as discussed, the outer (global) solutions for the hot ion mode plasma. 

It is worth mentioning, however, that there is the potential for additional synergies between rf-stablization and hot ion mode operation, in the case of inside-launch LHCD\cite{Ochs2015, Ochs2015b}. While $\alpha$-channeling can be used to redirect $\alpha$-particle power back in to the fuel-ion population using waves that deposit their energy on the ions, it might also be used for driving toroidal current by amplifying LHWs. It further happens to be the case that the favorable population inversions required for $\alpha$-channeling, inside launch LH penetration, and magnetic islands requiring stabilization all roughly coincide away from the core. This opens the possibility that, in a hot ion mode, rf-stabilization may not only be enhanced from the additional heat source entering the condensation effect as studied here, but may additionally benefit from LHWs that are amplified via $\alpha$-channeling. We raise this speculation here, but leave a dedicated analysis to future work.

In summary, we have demonstrated a novel benefit of hot ion mode operation: the enhanced stabilization of tearing modes via the rf condensation effect. Modest ion temperature surpluses and coupling strengths can significantly amplify a self-healing effect in which the rf-induced electron temperature perturbation focuses the ambient LHCD the to island center, thus stabilizing the island without the need for external aiming. Hot ion mode operation allows this passive stabilization to be achieved at smaller island widths, and with smaller RF powers. A preliminary investigation of the power-balance issues arising in the simultaneous maintenance of an ion temperature surplus and sufficient coupling confirms the accessibility of a strongly synergistic regime.

\begin{acknowledgements}
This work was supported by U.S. DOE Grants No. DE-AC02-09CH11466 and No. DE-SC0016072.
\end{acknowledgements}
\nocite{*}

\bibliography{aipsamp}

\end{document}